# Near-Field Acoustic Resonance Scattering of a Finite Bessel Beam by an Elastic Sphere

F.G. Mitri, *Member IEEE*

*Abstract* – The near-field acoustic scattering from a sphere centered on the axis of a *finite* Bessel acoustic beam is derived stemming from the Rayleigh-Sommerfeld diffraction surface integral and the addition theorems for the spherical wave and Legendre functions. The beam emerges from a finite circular disk vibrating according to one of its radial modes corresponding to the fundamental solution of a Bessel beam $J_0$. The incident pressure field's expression is derived analytically as a partial-wave series expansion taking into account the finite size and the distance from the center of the disk transducer. Initially, the scattered pressure by a rigid sphere is evaluated, and backscattering pressure moduli plots as well as 3-D directivity patterns for an elastic PMMA sphere centered on a finite Bessel beam with appropriate tuning of its half-cone angle, reveal possible resonance suppression of the sphere only in the zone near the Bessel transducer. Moreover, the analysis is extended to derive the mean spatial incident and scattered pressures at the surface of a rigid circular receiver of infinitesimal thickness. The transducer, sphere and receiver are assumed to be coaxial. Some applications can result from the present analysis since all physically realizable Bessel beam sources radiate finite sound beams as opposed to waves of infinite extent.

## I. INTRODUCTION

BESSEL beams constitute a class of non-diffracting waves [1] that present some advantages over the conventional Gaussian waves, such as the capability of trapping particulate matter over an extended distance [2] or producing a pulling force toward the acoustical source [3-7], as well as the ability to reconstruct after encountering an obstruction [8, 9]. When generated from a finite transducer [10-13], Bessel beams become of limited-diffraction and were successfully used in medical imaging [14, 15] for ultrasound tomography [16], non-destructive imaging [17] and particle manipulation and entrapment [18, 19] to name a few applications.

When the beam encounters a particle along its path, the acoustical scattering phenomenon occurs [20], and this effect forms the basis of image formation of inclusions in imaging applications as well as the transfer of linear momentum and the induced acoustic radiation force phenomenon in particle manipulation and single beam acoustic tweezers [21, 22].

Corresponding author: F.G. Mitri (e-mail: fmitri@gmail.com).

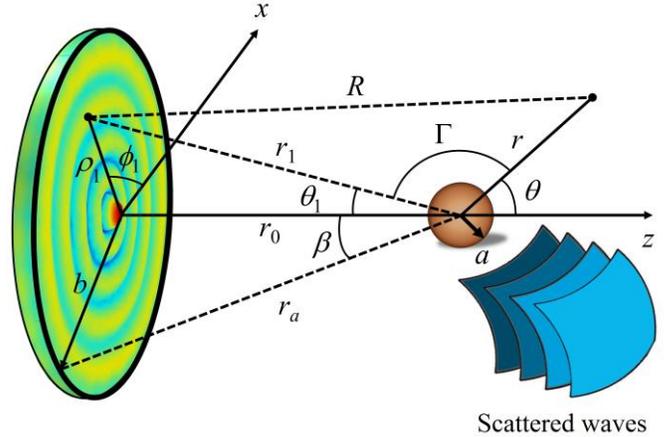

Fig. 1. Geometry of the problem used in the derivation for the scattering by a spherical target in the field of a finite zero-order Bessel beam.

To this end, the acoustic scattering from a spherical particle has only been examined for finite beams generated from circular plane [20, 23] and focused radiators [24, 25] with uniform vibration, and infinite Bessel beams [26-32]. However, considering the fact that practically every acoustic source (except a point source radiating omnidirectional waves) produces a finite beam, it is necessary to investigate the scattering of a Bessel beam generated from a finite aperture by a sphere.

The objective of this work is directed toward this aim, by deriving analytical equations for the incident and scattered waves of a finite zero-order Bessel beam incident upon an elastic sphere centered on its axis of wave propagation. Based on the Rayleigh surface integral [33] and the addition theorem for the spherical wave functions [34, 35] and Legendre functions, partial-wave series expansion are derived for the incident and the (resonance) scattered pressures. Moreover, a backscattering pressure is defined and numerical plots illustrate the theory with the display of the scattering directivity patterns in three-dimensions.

## II. METHOD

Consider a circular piston transducer surface of radius $b$, and a spherical target of radius $a$ centered on its axis (Fig. 1) and immersed in a non-viscous fluid. The sphere is situated at an axial distance $r_0$ from the surface of the circular transducer. The description of the incident acoustic field produced by the finite source, and represented by its scalar velocity potential field $\Phi_i$, is obtained from the Rayleigh-Sommerfeld integral as [33],

$$\Phi_i = \frac{1}{2\pi} \iint_{S_r} \frac{v e^{i(kR-\omega t)}}{R} dS_r, \quad (1)$$

where $R$ is the distance from the observation point to the finite source of circular surface $S_r$ (Fig. 1), and $v$ is the normal velocity at $z = 0$. In the case of a uniform vibrational surface, $v$ is unitary [36-40], but takes particular values when the radiator is simply supported [36, 37, 41], clamped [36, 37], or excited with a Gaussian apodization [37].

For the purpose of the present study, an apodization in the form of a cylindrical Bessel function of order zero ($J_0$) is considered [42, 43], which produces a Bessel beam with a maximum in amplitude at the center of the beam. It is suggested here that the fundamental Bessel beam (of order zero) can be generated from a finite piezoelectric transducer vibrating according to its radially symmetric modes (Fig. 2) that can be excited, although a non-uniform poling technique [10, 44, 45], or a multi-annular probe driven by a signal in the form of a Bessel function [14, 16] may be used for the same purpose. Adequate identification of the resonance vibrational modes and the deformed shaped of the disk required developing a 3D finite-element model using Comsol, Inc., via a solid mechanics eigen-frequency analysis. The following parameters for the circular disk were chosen; diameter = 45 mm, thickness of 3 mm, density of 7500 kg/m$^3$, a Poisson's ratio of 0.33, and a Young's modulus of 125 GPa. The disk was completely meshed using 102043 tetrahedral linear solid elements, 12070 triangular elements, 336 edge elements and 8 vertex elements through the "mesh" command. The vibrational modes and their associated resonance frequencies were calculated and the results displayed in Fig. 2. The study was only focused on the radially-symmetric modes, however, there exist other vibrational modes, known as the flexural (anti-symmetric) modes, that exhibit Bessel-like vibrational profiles. Nevertheless, those are in practice not related to the piezoelectricity of the vibrating disk and cannot be excited by the voltage across the disk surfaces [46]. One may use, however, a mechanical shaker/vibrator directly connected to the disk transducer and excite such modes to potentially produce a propagating beam.

Taking into account the geometry of the problem as shown in Fig. 1, the normal velocity is expressed as,

$$v|_{z=0} = V_0 J_0(k_\rho \rho_1), \quad (2)$$

where $V_0$ is the velocity amplitude, $k_\rho = k \sin \beta_m$, and $\rho_1$ is the distance from the center of the radiator to a point on its flat surface. The parameter $k$ is the wave number of the acoustic radiation, and $\beta_m$ is defined as the half-cone angle of the beam that can be directly connected with each radial resonance vibrational mode $m$, which are shown in Fig. 2.

Suppressing the time dependence $e^{-i\omega t}$ for convenience, and using the addition theorem for the spherical functions, i.e. (10.1.45) and (10.1.46) in [35] such that $r \leq r_1$, (1) is expressed as,

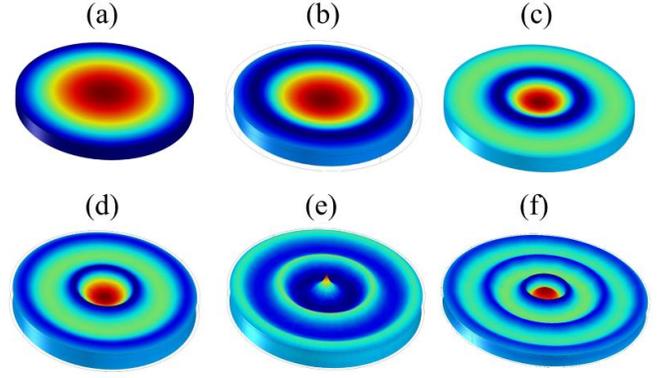

Fig. 2. Panels (a)-(f) display the first six radially-symmetric resonance modes of a piezoelectric (Lead Zirconate Titanate PZT-4) crystal corresponding to Bessel-type vibrations with variable half-cone angles. See also the corresponding animations for each of the modes. The resonance frequencies, obtained using a finite-element analysis using Comsol, Inc., are listed as follows; (a) 63 kHz, (b) 164 kHz, (c) 258.5 kHz, (d) 348 kHz, (e) 430.5 kHz, and (f) 502 kHz.

$$\Phi_i = \frac{ikV_0}{2\pi} \sum_{n=0}^{\infty} (2n+1) j_n(kr) \\ \times \iint_{S_r} h_n^{(1)}(kr_1) J_0(k_\rho \rho_1) P_n(\cos \Gamma) dS_r, \quad (3)$$

where $j_n(\cdot)$ and $h_n^{(1)}(\cdot)$ are the spherical Bessel and Hankel functions of the first kind, $P_n(\cdot)$ are the Legendre functions, and the differential surface $dS_r = \rho_1 d\rho_1 d\phi_1 = r_1 dr_1 d\phi_1$, since $r_1^2 = \rho_1^2 + r_0^2$.

Making use of the addition theorem for the Legendre functions [47, 48] using the definition of the angles as given in Fig. 1, $P_n(\cos \Gamma)$ can be expressed as ((3.19), p. 65 in [49]),

$$P_n(\cos \Gamma) = \sum_{\ell=0}^{n} (2 - \delta_{\ell,0}) \frac{(n-\ell)!}{(n+\ell)!} \\ \times P_n^\ell(\cos(\pi-\theta)) P_n^\ell(\cos\theta_1) \cos\ell(\phi-\phi_1), \quad (4)$$

where $\delta$ is the Kronecker delta function, and $P_n^\ell(\cdot)$ are the associated Legendre functions. Integrating both sides of (4) with respect to $\phi_1$ gives,

$$\int_0^{2\pi} P_n(\cos \Gamma) d\phi_1 = 2\pi(-1)^n P_n(\cos\theta) P_n(\cos\theta_1). \quad (5)$$

Substituting (5) into (3), and after arithmetic manipulation, the incident velocity potential can be expressed as,

$$\Phi_i = \Phi_0 \sum_{n=0}^{\infty} \Lambda_{J_0,n} i^n (2n+1) j_n(kr) P_n(\cos\theta), \quad (6)$$

where,

$$\Lambda_{J_0,n} = i^n \int_{kr_0}^{kr_a} (kr_1) h_n^{(1)}(kr_1) J_0\left(k_\rho \sqrt{r_1^2 - r_0^2}\right) P_n\left(\frac{r_0}{r_1}\right) d(kr_1), \quad (7)$$

and $\Phi_0 = iV_0/k$.

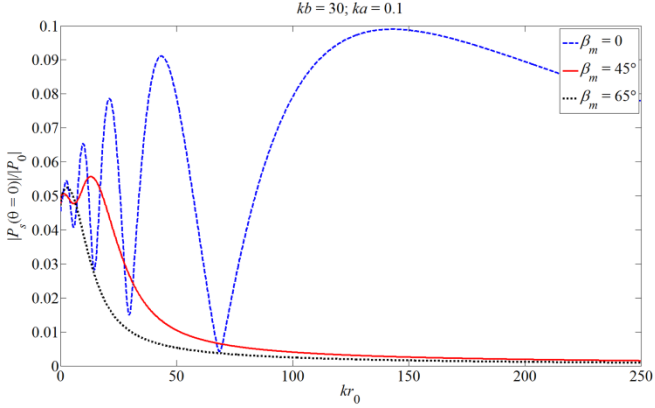

Fig. 3. The on-axis forward scattering pressure magnitude from a rigid immovable sphere at $r = a$ in the Rayleigh regime ($ka = 0.1$) with $kb = 30$. The uniformly vibrating rigid piston result corresponds to the dashed curve ($\beta_m = 0$). Note that the maximum in the dashed plot occurs at $kr_0 = (kb)^2/(2\pi) = 144$. The solid and dotted plots correspond to finite zero-order Bessel beams for $\beta_m = 45°$ and $65°$, respectively.

For the case where $\beta_m = 0$ and $kr_a$ (or $kb$) $\to \infty$, (7) reduces to,

$$\Lambda_{J_0,n} = \Lambda_n = e^{ikr_0}, \quad (8)$$

because any real medium is more or less absorptive (i.e. $k$ becomes complex) and the term $e^{ikr_a}$ resulting from the asymptotic form of the spherical Hankel function of the first kind will vanish. This case corresponds to infinite plane progressive waves, and an equivalent form was given previously assuming a time dependence in the form of $e^{i\omega t}$ [38, 39].

Using the simplest case of the Mehler-Heine formula [35, $\mu = 0$ in (9.1.71)], it is particularly interesting to note from (7) that the function $J_0\left(k_\rho\sqrt{r_1^2 - r_0^2}\right) \approx P_n\left(\cos\beta_m\right)$, provided that $\sqrt{r_1^2 - r_0^2} \sim (n+1/2)/k$, and in the limit where the partial-wave number $n \to \infty$, and the half-cone angle $\beta_m \to 0°$ [27]. In this limit, (7) can be expressed as,

$$\Lambda_{J_0,n}\Big|_{\substack{n \gg 1 \\ \beta_m \ll 57.3°}} = i^n P_n\left(\cos\beta_m\right) f_n, \quad (9)$$

where [21, 38-40],

$$f_{n\geq 2} = -f_{n-2} + (kr_a) h^{(1)}_{n-1}(kr_a)\left[P_{n-2}\left(\frac{r_0}{r_a}\right) - P_n\left(\frac{r_0}{r_a}\right)\right], \quad (10)$$

with $f_0 = e^{ikr_0} - e^{ikr_a}$, and $f_1 = \left[e^{ikr_0} - (r_0/r_a)e^{ikr_a}\right]/i$.

It is obvious from (9) that the plane wave limit can be immediately recovered for $\beta_m = 0$. Moreover, in the particular case of a radiator vibrating back and forth like a rigid piston, the function $f_n$ has an exact solution [21, 38-40] as given by (10). However, beyond the limit $n \gg 1$, and $\beta_m \ll 57.3°$, it is not yet established if (7) in its present form will have a closed form solution though various tests have been performed without a conclusive result. Therefore, for the purpose of the present study, (7) is evaluated numerically using the trapezoidal method.

Note also that on the axis of wave propagation ($r = 0$), only the leading term (given after (10)) $f_0 \neq 0$. Therefore, along the axis, (7) can be further simplified and is expressed as,

$$\Lambda_{J_0,0} = -i\int_{kr_0}^{kr_a} J_0\left(k_\rho\sqrt{r_1^2 - r_0^2}\right)e^{ikr_1} d(kr_1), \quad (11)$$

then in this case (6) reduces to $\Phi_i^{axial} = \Phi_0 \Lambda_{J_0,0}$.

The incident pressure field can also be evaluated as,

$$\begin{aligned} P_i &= i\omega\rho_0 \Phi_i, \\ &= P_0 \sum_{n=0}^{\infty} \Lambda_{J_0,n} i^n (2n+1) j_n(kr) P_n(\cos\theta), \end{aligned} \quad (12)$$

where $P_0 = \rho_0 c_0 |V_0|$, is the pressure amplitude, and $\rho_0$ and $c_0$ the density and speed of sound in the fluid medium in the absence of acoustic disturbance.

The presence of the sphere in the beam's path produces a scattered wave, for which the scattered pressure is expressed as,

$$P_s = P_0 \sum_{n=0}^{\infty} \Lambda_{J_0,n} i^n (2n+1) S_n h_n^{(1)}(kr) P_n(\cos\theta), \quad (13)$$

where $S_n$ are the scattering coefficients to be determined from appropriate boundary conditions. For an elastic sphere [50], the normal component of stress in the elastic solid at the interface equals the pressure in the fluid, the normal (radial) component of displacement (or velocity, respectively) of the solid at the interface must equal the normal component of displacement (or velocity, respectively) of the fluid, and the tangential components of shearing stress must vanish at the surface of the solid (since the exterior fluid medium is considered non-viscous). Rewriting the boundary conditions, the scattering coefficients are found to equal those obtained from the study of acoustic scattering of infinite plane progressive waves [50]. They are expressed as,

$$S_n = \begin{vmatrix} A_1^* & d_{12} & d_{13} \\ A_2^* & d_{22} & d_{23} \\ A_3^* & d_{32} & d_{33} \end{vmatrix} \Bigg/ \begin{vmatrix} d_{11} & d_{12} & d_{13} \\ d_{21} & d_{22} & d_{23} \\ d_{31} & d_{32} & d_{33} \end{vmatrix}, \quad (14)$$

where the dimensionless matrix elements $A_i^*$ and $d_{ij}$ are given explicitly in the Appendix of [50].

The formalism of the Resonance Scattering Theory (RST) developed previously for plane waves [51], and later extended to any types of beams [52] reveals that each individual normal-mode contribution to the total scattering pressure amplitude consists of a background term which is a smooth function of frequency. Superimposed upon this background (and interfering with it) is a series of resonances, which closely coincide with the eigen-frequencies of vibration of the fluid-loaded elastic scatterer. The present formalism can be extended to account for the resonances of the elastic sphere by defining a "resonance" scattering pressure for which an appropriate background term (rigid, soft or intermediate) is subtracted from the "total" scattered pressure field (13). From the RST formalism, it has been found that for dense solid spheres, the background is closely approximated by the modal amplitudes of rigid-body scattering, while for a very soft sphere it approaches the soft-body scattering amplitude. However, for spheres with density comparable to the external fluid, the intermediate background has been adequately introduced to properly identify the resonances [53]. The scattering coefficients for rigid immovable and soft bodies are given in standard texts [51].

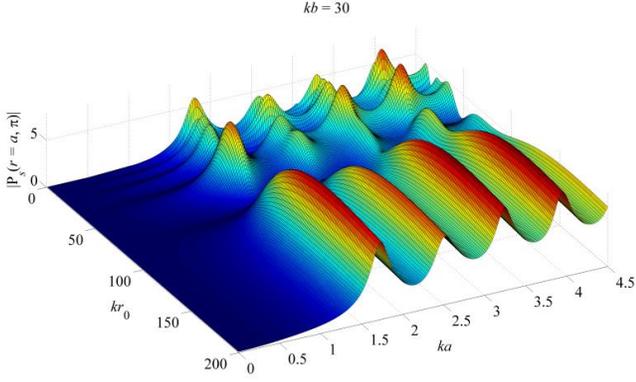

Fig. 4. Backscattering pressure modulus for a PMMA elastic sphere in water with a circular disk with uniform vibration (i.e. $\beta_m = 0$) as a function of $ka$, $kr_0$ at a fixed dimensionless radius $kb = 30$. Note the quadrupole ($n = 2$) resonance at $ka \sim 1.74$, the hexapole ($n = 3$) resonance at $ka \sim 2.53$, the octupole ($n = 4$) resonance at $ka \sim 3.26$ (and some residual structure in the scattered pressure from the dipole resonance $n = 1$), and the decapole ($n = 5$) resonance at $ka \sim 3.96$.

The resonance scattering pressure can be therefore expressed as,

$$P_s^{res} = P_0 \sum_{n=0}^{\infty} \Lambda_{J_0,n} i^n (2n+1)\left[ S_n - S_n^{(r,s,i)} \right] h_n^{(1)}(kr) P_n(\cos\theta), \quad (15)$$

where $S_n^{(r,s,i)}$ are the scattering coefficients corresponding to a rigid, soft, or intermediate backgrounds, respectively.

### III. NUMERICAL RESULTS AND DISCUSSION

The analysis is started by verifying the accuracy of the numerical integration (trapezoidal) method with a closed-from solution for a known beam. Assuming a uniform vibration of a circular piston transducer, the expression for the parameter

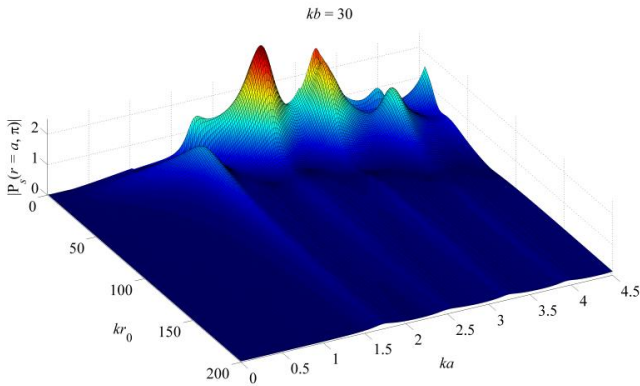

Fig. 5. Backscattering pressure modulus for a PMMA elastic sphere in water with a finite Bessel beam with $\beta_{m,2} = 54.7346°$, as a function of $ka$ and $kr_0$ with $kb = 30$. The resonance associated with the quadrupole mode with $n = 2$ near $ka \sim 1.74$ is partially suppressed, and it is still weakly manifested as $kr_0$ increases.

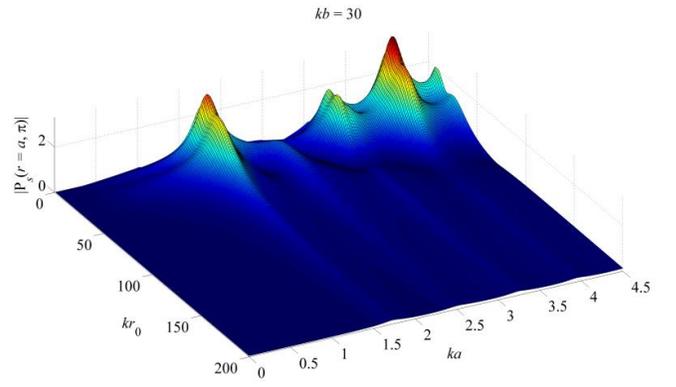

Fig. 6. Backscattering pressure modulus for a PMMA elastic sphere in water with a finite Bessel beam with $\beta_{m,3} = 39.2315°$, as a function of $ka$ and $kr_0$ with $kb = 30$. The resonance associated with the hexapole mode with $n = 3$ near $ka \sim 2.53$ is weakly manifested as $kr_0$ increases.

$\Lambda_{J_0,n}$ given by (7) is reduced to [38, 39],

$$\Lambda_{J_0,n} = \Lambda_n = i^n f_n, \quad (16)$$

where $f_n$ is given by (10). To validate (7), the exact solution (10) is computed and compared with the results of the numerical integration with a chosen sampling step of $\delta(kr_1) = 10^{-4}$. With this parameter, it has been found that the absolute error for a broad range of chosen sets $(kr_0, kr_b)$ is below $10^{-7}$. Thus, a smaller sampling step is not required since the absolute error can be safely considered as negligible.

The analysis is started by analyzing the beam profile through the forward scattering ($\theta = 0$) pressure versus $kr_0$ as given by (13) at $r = a$, from a rigid immovable sphere in the Rayleigh limit (i.e. $ka = 0.1$). This can closely simulate an experimental beam-profile measurement using a needle hydrophone. For the rigid immovable sphere, the scattering coefficients are expressed as $S_n^r = -j_n'(ka)/h_n^{(1)'}(ka)$ [51]. The first case corresponds to a finite radiator with uniform vibration (i.e. $\beta_m = 0$), whereas the second and third ones correspond to a finite radiator with a vibration profile described by a cylindrical Bessel function $J_0$ with $\beta_m = 45°$ and $65°$, respectively. For the sake of the present analysis, the dimensionless radius of the radiating source is selected to be $kb = 30$, but any other finite value may be chosen. If $kr_0 < (kb)^2/(2\pi)$, the sphere is situated in the near-field (Fresnel) zone of the transducer assuming a uniform vibration Otherwise, it is placed in the far-field (Fraunhofer) region of the transducer when $kr_0 \geq (kb)^2/(2\pi)$ (p. 165 in [54]). (Note that these near (Fresnel)- and far-field (Fraunhofer) definitions only hold for the piston transducer vibrating back and forth with constant normal velocity of its surface, and may not hold for the radial or other anti-symmetric vibrational modes where the vibration velocity is not constant over its surface). For the uniform vibration case (i.e. $\beta_m = 0$), the rapid oscillations in the near-field zone determined by $kr_0 < 144$ (i.e. dashed curve in Fig. 3) are the result of the forward scattering that is affected by the incident diffracted field from the finite aperture vibrating back and forth in a rigid piston mode. This is not the case when the radiator vibrating velocity takes the form of a cylindrical Bessel function $J_0$. The solid and dotted lines in Fig. 3 show that the modulus of the forward scattering pressure decays rapidly in the region defined by $kr_0 < 50$, however, it has minimal variations beyond that limit. Moreover, as $\beta_m$ increases, the forward scattering is reduced (dotted curve).

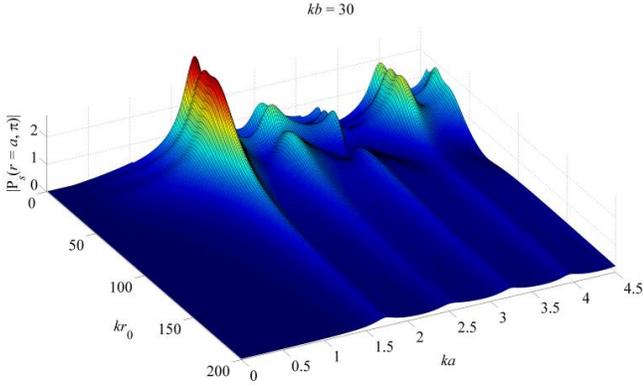

Fig. 7. Backscattering pressure modulus for a PMMA elastic sphere in water with a finite Bessel beam with $\beta_{m,4} = 30.5556°$, as a function of $ka$ and $kr_0$ with $kb = 30$. The resonance associated with the octupole mode with $n = 4$ near $ka \sim 3.26$ is totally suppressed around $kr_0 \sim 31$ (See also Fig. 10), however, it is weakly manifested because of the residual contribution of the dipole mode ($n = 1$).

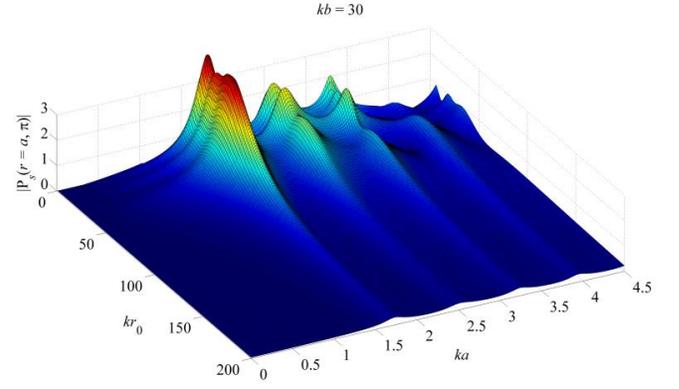

Fig. 8. Backscattering pressure modulus for a PMMA elastic sphere in water with a finite Bessel beam with $\beta_{m,5} = 25.0173°$, as a function of $ka$ and $kr_0$ with $kb = 30$. The resonance associated with the decapole mode with $n = 5$ near $ka \sim 3.96$ is partially suppressed only in the near-field of the transducer, however, it is still weakly manifested as $kr_0$ increases.

Now, the case of an elastic sphere immersed in water ($\rho_0 = 1000$ kg/m$^3$, $c_0 = 1479$ m/s) is considered. The sphere is assumed to be made of a polymethylmetacrylate (PMMA) *elastic* material having the following properties, $\rho_s = 1190$ kg/m$^3$, $c_L = 2690$ m/s, $c_S = 1340$ m/s. The modulus of the backscattering ($\theta = \pi$) pressure (13) is computed at $r = a$, with particular emphasis on the values of the half-cone angle $\beta$, the dimensionless frequency $ka$, as well as the dimensionless distance from the source to the center of the sphere $kr_0$ for a dimensionless radius of the circular transducer $kb = 30$.

For a disk vibrating uniformly, the half-cone angle $\beta_m$ is set to zero so as the normal velocity given by (2) equals unity. In this configuration, the modulus of the backscattering pressure is displayed in Fig. 4. In the Fresnel zone ($kr_0 < 144$), the backscattering pressure undergoes oscillations, which follow the variations of the incident beam in the near-field of the transducer [36, 37]. The maximum in the $\left|P_s(r=a,\pi)\right|$ plot is reached at $kr_0 = (kb)^2/(2\pi) = 144$, which shows a smooth decay in the far-field zone $kr_0 \geq 144$. Moreover, all four resonances obtained in the infinite plane wave case (solid curve, Fig. 2 in [27]) have been clearly detected in the $\left|P_s(r=a,\pi)\right|$ plot, both in the near- and far-field zones of the transducer, but with different amplitudes.

The modifications in the backscattering from the finite Bessel beam illumination with different half-cone angle values are now investigated. Earlier works (assuming infinite beams) showed that resonance suppression can be achieved [27, 55, 56] by fine-tuning the half-cone angle $\beta_m$ to correspond to the roots of the Legendre function $P_n(\cos\beta_m) = 0$. The 6-digit approximations to $\beta_{m,n}$ for suppressing resonance modes with their corresponding partial-waves $n = 2, 3, 4,$ and $5$, are $\beta_{m,n} =$ 54.7346°, 39.2315°, 30.5556°, and 25.0173°, respectively [27]. Those angles are used here to compute the backscattering pressure modulus for a PMMA elastic sphere in water as a function of $ka$, $kr_0$ at a fixed dimensionless radius $kb = 30$. Figs. 5-8 show the corresponding plots of the backscattered pressure field modulus at those half-cone angles, respectively. The inspection of each of those plots shows that each particular resonance may be indeed altered but not totally suppressed especially in the zone near the transducer.

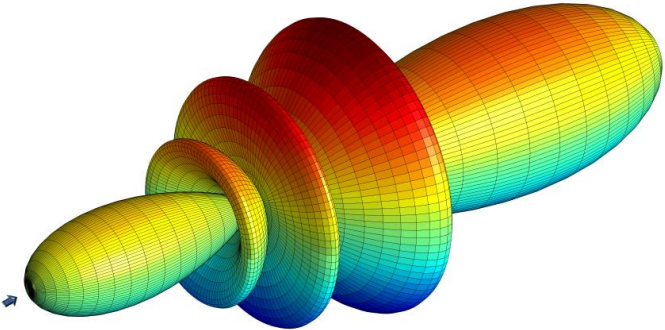

Fig. 9. The 3-D directivity pattern of the scattering pressure modulus for a PMMA elastic sphere in water at $ka = 3.26$, $kr_0 = 31$, $kb = 30$ and $\beta_m = 0$. Notice the backscattering lobe in the direction $\theta = \pi$ corresponding to the octupole ($n = 4$) resonance mode. The arrow on the left-hand side of the panel indicates the direction of the incident waves generated from a finite circular piston transducer.

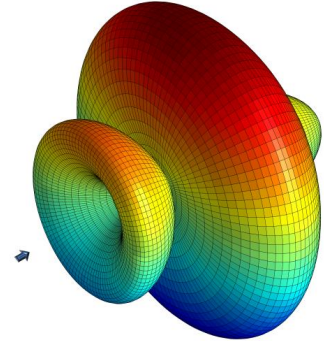

Fig. 10. The same as in Fig. 9 but the half-cone angle $\beta_{m,4} = 30.5556°$. The backscattering lobe in the direction $\theta = \pi$ appearing in Fig. 9 is now totally suppressed as the half-cone angle of the zeroth-order Bessel beam is appropriately tuned to cancel the backscattering. The arrow on the left-hand side of the panel indicates the direction of the incident finite Bessel beam.

This is also clearly displayed in Figs. 9,10, in which the 3-D directivity pattern of the far-field scattering form function modulus are plotted for $\beta_m = 0$ (no resonance suppression – Fig. 9) and for $\beta_{m,4} = 30.5556°$ (suppression of the octupole ($n = 4$) resonance using a finite $J_0$ Bessel beam – Fig. 10) at $ka = 3.26$, $kr_0 = 31$, $kb = 30$. Notice that the backscattering lobe in the backward direction $\theta = \pi$ (Fig. 9) is completely suppressed (Fig. 10) in the zone near the Bessel radiator ($kr_0 = 31$).

On the other hand, Figs. 5-8 show that the resonances in the backscattering pressure plots are still manifested (though weakly) in the zone far from transducer for large $kr_0$. The computational plots clearly display the resulting changes due to the effects of diffraction of the finite Bessel beam as opposed to the ideal case of waves of infinite extent [27]. Precisely, a complete suppression of the resonances may occur only in the zone near the Bessel transducer (due to diffraction effects); however, unlike the infinite Bessel beam case, those resonance vibrational features in the scattering remain manifested, though weakly, in the zone far from the transducer.

It is also important to note that (12) and (13) are applicable to determine the incident pressure field generated from the finite radiator, and the scattered pressure field from the sphere at an infinitesimal point in space. However, an acoustic receiver of finite area is often used to detect the emitted waves from the vibrating Bessel source and/or the scattered waves from the sphere. In those circumstances, the output of the receiver becomes proportional to the spatially-averaged acoustical pressure acting on its surface [33, 39, 41, 57]. Thus, (12) and (13) have to be integrated over the receiver's surface to obtain the mean-spatial pressures. For the reader's convenience, the expressions for the mean-spatial pressures are derived and presented in the Appendix. Their computations and further evaluation may be used to advantage for experimental design purposes, which go beyond the scope of the present research.

In this analysis, the aim was to investigate the near-field ($r \leq r_1$, $r \leq r_a$) resonance scattering and possible resonance alteration/suppression of an elastic sphere immersed in a non-viscous fluid. Yet, the analysis can be readily extended to investigate the far-field scattering field from the sphere ($r \geq r_1$, $r \geq r_a$) by modification of the theory according to Ref. [39]. Moreover, the sphere was assumed to be centered on the axis of wave propagation of a finite Bessel beam. There exist, however, cases where the sphere can be placed arbitrarily in the acoustic field [58, 59], so that the symmetry in the scattering is broken [32, 52]. The present analysis can be further extended to that particular case provided that the appropriate beam-shape coefficients [58] are used along with the scattering coefficients of the particle with any geometry and material type (fluid/elastic/ viscoelastic sphere, shell with a hollow, multilayered sphere/shell etc.). Moreover, the scattering phenomenon is of particular importance in imaging applications using Bessel beams [60], as well as the evaluation/computation of the 3D components of the acoustic radiation force [7, 61] and torque [62, 63] in the development of novel acoustical tweezers devices, and this analysis can be helpful along this line of research. In addition, the theory can be extended to account for the viscosity of the surrounding fluid medium around the spherical target, which can be of some relevance in applications involving highly viscous [64] or thermoviscous [65, 66] fluids.

## IV. CONCLUSION

In summary, the near-field resonance scattering theory from an elastic sphere immersed in a non-viscous fluid and centered on the axis of a finite Bessel beam is investigated. The moduli of the backscattering pressure are evaluated for various situations where the incident waves originate from a uniform vibration, or a Bessel beam excitation of the transducer's surface. Moreover, the 3-D directivity patterns at the surface of the sphere demonstrate that complete suppression of the resonances may occur only in the zone near the Bessel transducer; however, those remain manifested, though weakly, in the zone far from the transducer, mainly because of diffraction effects. In addition, the mean-spatial incident and scattered pressures on the surface of a rigid immovable receiver are derived for the case where the finite acoustic Bessel source, sphere and receiver are aligned coaxially. The present analysis can be applied to study the scattering from any type and size of spherical particles, and can be extended to account for the shift if the sphere is arbitrarily located in the field of a finite Bessel beam.

## APPENDIX

In this Appendix, mathematical expressions using partial-wave expansions, representing the mean-spatial pressures are provided. For the incoming waves, the mean-spatial incident pressure is expressed as,

$$\overline{P_i} = \frac{1}{S_2} \iint_{S_2} P_i \, dS_2, \tag{17}$$

where $S_2$ is the receiver's surface, the over-bar denotes spatial averaging, and the differential surface $dS_2 = \rho_2 d\rho_2 d\phi_2 = r dr d\phi_2$, since $r^2 = \rho_2^2 + (r_2 - r_0)^2$ (See Fig. 11). Substituting (12) into (17) and manipulating the result, the *mean-spatial* incident pressure is expressed as,

$$\overline{P_i} = \frac{2P_0}{(kb_2)^2} \sum_{n=0}^{\infty} \Lambda_{J_0, n} i^n (2n+1) \int_{k(r_2-r_0)}^{kr_b} (kr) j_n(kr) P_n\left(\frac{(r_2 - r_0)}{r}\right) d(kr),$$

$$r_b = \sqrt{b_2^2 + (r_2 - r_0)^2} \quad (18)$$

where $r_b = \sqrt{b_2^2 + (r_2 - r_0)^2}$ (Fig. 11).

It is important to note that the definite integral,

$$f_n^i = \int_{k(r_2-r_0)}^{kr_b} (kr) j_n(kr) P_n\left(\frac{(r_2-r_0)}{r}\right) d(kr), \quad (19)$$

has a closed-form solution [38, 39], which is expressed as,

$$f_{n\geq 2}^i = -f_{n-2}^i + (kr_b) j_{n-1}(kr_b)\left[P_{n-2}\left(\frac{r_2-r_0}{r_b}\right) - P_n\left(\frac{r_2-r_0}{r_b}\right)\right], \quad (20)$$

with $f_0^i = \cos[k(r_2-r_0)] - \cos(kr_b)$, and

$$f_1^i = \sin[k(r_2-r_0)] - [(r_2-r_0)/r_b]\sin(kr_b).$$

Thus, using (20), (18) can be expressed as,

$$\overline{P_i} = \frac{2P_0}{(kb_2)^2} \sum_{n=0}^{\infty} f_n^i \Lambda_{J_0,n} i^n (2n+1). \quad (21)$$

For the scattered waves, a mean-spatial scattered pressure can be defined as,

$$\overline{P_s} = \frac{1}{S_2} \iint_{S_2} P_s \, dS_2. \quad (22)$$

Substituting (13) into (22) and manipulating the result, the *mean-spatial* scattered pressure at the receiver is expressed as,

$$\overline{P_s} = \frac{2P_0}{(kb_2)^2} \sum_{n=0}^{\infty} f_n^s \Lambda_{J_0,n} i^n (2n+1) S_n, \quad (23)$$

where

$$f_n^s = \int_{k(r_2-r_0)}^{kr_b} (kr) h_n^{(1)}(kr) P_n\left(\frac{(r_2-r_0)}{r}\right) d(kr), \quad (24)$$

has also a closed-form solution [38, 39], which is expressed as,

$$f_{n\geq 2}^s = -f_{n-2}^s + (kr_b) h_{n-1}^{(1)}(kr_b)\left[P_{n-2}\left(\frac{r_2-r_0}{r_b}\right) - P_n\left(\frac{r_2-r_0}{r_b}\right)\right], \quad (25)$$

with $f_0^s = e^{ik(r_2-r_0)} - e^{ikr_b}$, and

$$f_1^s = \left[e^{ik(r_2-r_0)} - \frac{(r_2-r_0)}{r_b} e^{ikr_b}\right]/i.$$

Similarly to (15) and (23), one can define a mean-spatial resonance scattered pressure as

$$\overline{P_s^{res}} = \frac{2P_0}{(kb_2)^2} \sum_{n=0}^{\infty} f_n^s \Lambda_{J_0,n} i^n (2n+1)\left[S_n - S_n^{(r,s,i)}\right]. \quad (26)$$

where $S_n^{(r,s,i)}$ are the scattering coefficients corresponding to a rigid, soft, or intermediate backgrounds, respectively.

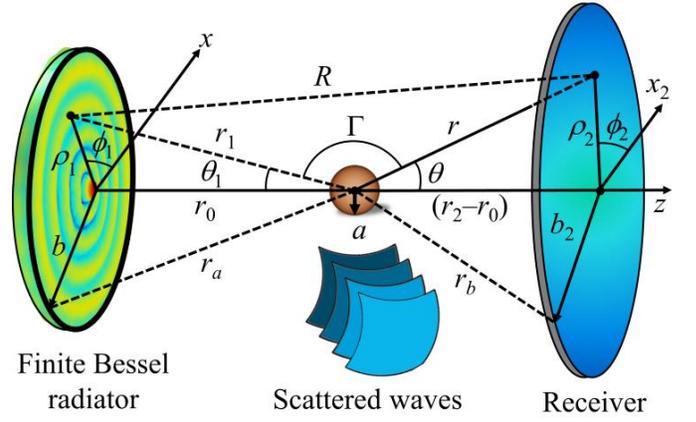

Fig. 11. Geometry of the problem. A circular receiver is used to detect either the mean spatial incident pressure from the source (in the absence of the sphere), or the scattered waves from the sphere. The distances radiator-sphere and sphere-receiver are denoted by $r_0$ and $(r_2-r_0)$, respectively.